\documentclass[conference]{IEEEtran}
\IEEEoverridecommandlockouts
\usepackage{cite}
\usepackage{amsmath,amssymb,amsfonts}
\usepackage{algorithmic}
\usepackage{graphicx}
\usepackage{comment}
\usepackage{textcomp}
\usepackage[table]{xcolor}
\def\BibTeX{{\rm B\kern-.05em{\sc i\kern-.025em b}\kern-.08em
    T\kern-.1667em\lower.7ex\hbox{E}\kern-.125emX}}
\usepackage{orcidlink}    

\usepackage{booktabs,multirow,bigstrut}
\usepackage[nameinlink,noabbrev,capitalise]{cleveref}

\hypersetup{hidelinks}
\usepackage{textgreek}
\begin{document}

\title{A Low-Power Spike Detector Using In-Memory Computing for Event-based Neural Frontend
\thanks{The work done in this paper was partially supported by a grant from the Research Grants Council of the Hong Kong Special Administrative Region, China (Project No. CityU 11200922).}
}
\author{ Ye~Ke~\orcidlink{0009-0002-9809-1192}, and Arindam~Basu~\orcidlink{0000-0003-1035-8770} \\
\textit{Department of Electrical Engineering, City University of Hong Kong}\\
Email: yeke22-c@my.cityu.edu.hk, arinbasu@cityu.edu.hk
}

\maketitle

\begin{abstract}
With the sensor scaling of next-generation Brain-Machine Interface (BMI) systems, the massive A/D conversion and analog multiplexing at the neural frontend poses a challenge in terms of power and data rates for wireless and implantable BMIs. 
While previous works have reported the neuromorphic compression of neural signal, further compression requires integration of spike detectors on chip. In this work, we propose an efficient HRAM-based spike detector using In-memory computing for compressive event-based neural frontend. Our proposed method involves detecting spikes from event pulses without reconstructing the signal and uses a 10T hybrid in-memory computing bitcell for the accumulation and thresholding operations. We show that our method ensures a spike detection accuracy of 92-99\% for neural signal inputs while consuming only $13.8$ nW per channel in $65$ nm CMOS.

\end{abstract}

\begin{IEEEkeywords}
Brain-machine interfaces (BMI), neuromorphic compression, spike detection, in-memory computing (IMC).
\end{IEEEkeywords}

\section{Introduction}
Brain-machine interfaces (BMI) opened up the possibility of real-time control of prosthetic devices for patients with paralysis and other neurological diseases. In this scenario, the neural activities are recorded by the micro-electrode array (MEA) to analyze the neural spikes (action potentials or AP) for further intention decoding. Next-generation BMI systems are expected to support the parallel recording of thousands of channels to improve decoding performance and enable complex control of effectors\cite{zhang_algorithm_2022} \cite{even-chen_power-saving_2020}. However, increasing the number of channels, $N_{chan}$, have brought challenges in digitizing the immense amount of neural data and transmitting it off-chip within the power and bandwidth constraints of wireless distributed BMI systems. Additionally, the use of massive analog multiplexing can decrease the accuracy and efficiency of the neural frontend.

\begin{figure}[t!]
\centering
\resizebox{0.5\textwidth}{!}{
\includegraphics[width=\textwidth]{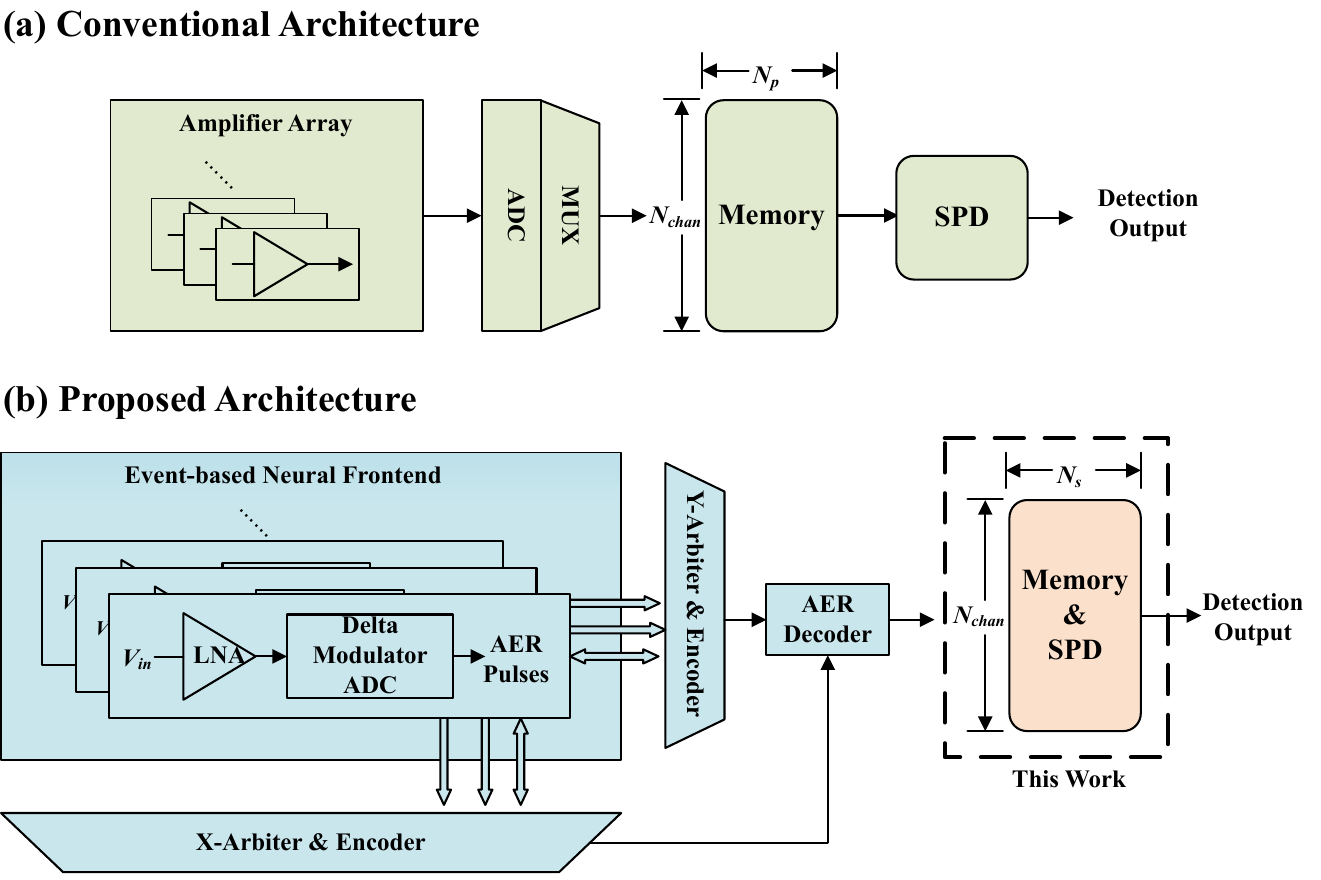}
}
\caption{Comparing system architecture of (a) conventional and (b) proposed HRAM-based IMC-SPD for event-based neural frontend. The compressive neural frontend consists of a low-noise amplifier and a variable gain amplifier combined with a delta modulator in each channel, similar to the architecture in \cite{mohan_architectural_2023}.}
\label{fig:Fig1}
\end{figure}

Various data compression efforts have been made to minimize digitization and multiplexing at the neural frontend and realize scalable, low-power, and wireless BMI systems. Inspired by event-based image sensors, \cite{muratore_data-compressive_2019} and \cite{yan_data_2022} exploited the spatial and temporal sparsity of neural signals through wired-OR interactions within a single-slope ADC array. A lossy compression was achieved by dropping samples around the baseline while retaining critical samples that indicate the neural spikes. An event-based neural recording (EBNR) system proposed in \cite{corradi_neuromorphic_2015} utilized address event representation (AER) to convert the neural signals into asynchronous digital pulse streams through in-pixel thresholding and thus digitize data mostly during action potentials. \cite{mohan_architectural_2023} extended \cite{corradi_neuromorphic_2015} to assess the potential for scaling EBNR to thousands of channels with collision management. While EBNR is promising, further data reduction is needed for the scalability of this approach since data rates remain higher than typical wireless transmission capabilities.

\begin{figure*}[t!]
\centering
\resizebox{\textwidth}{!}{
\includegraphics[width=\textwidth]{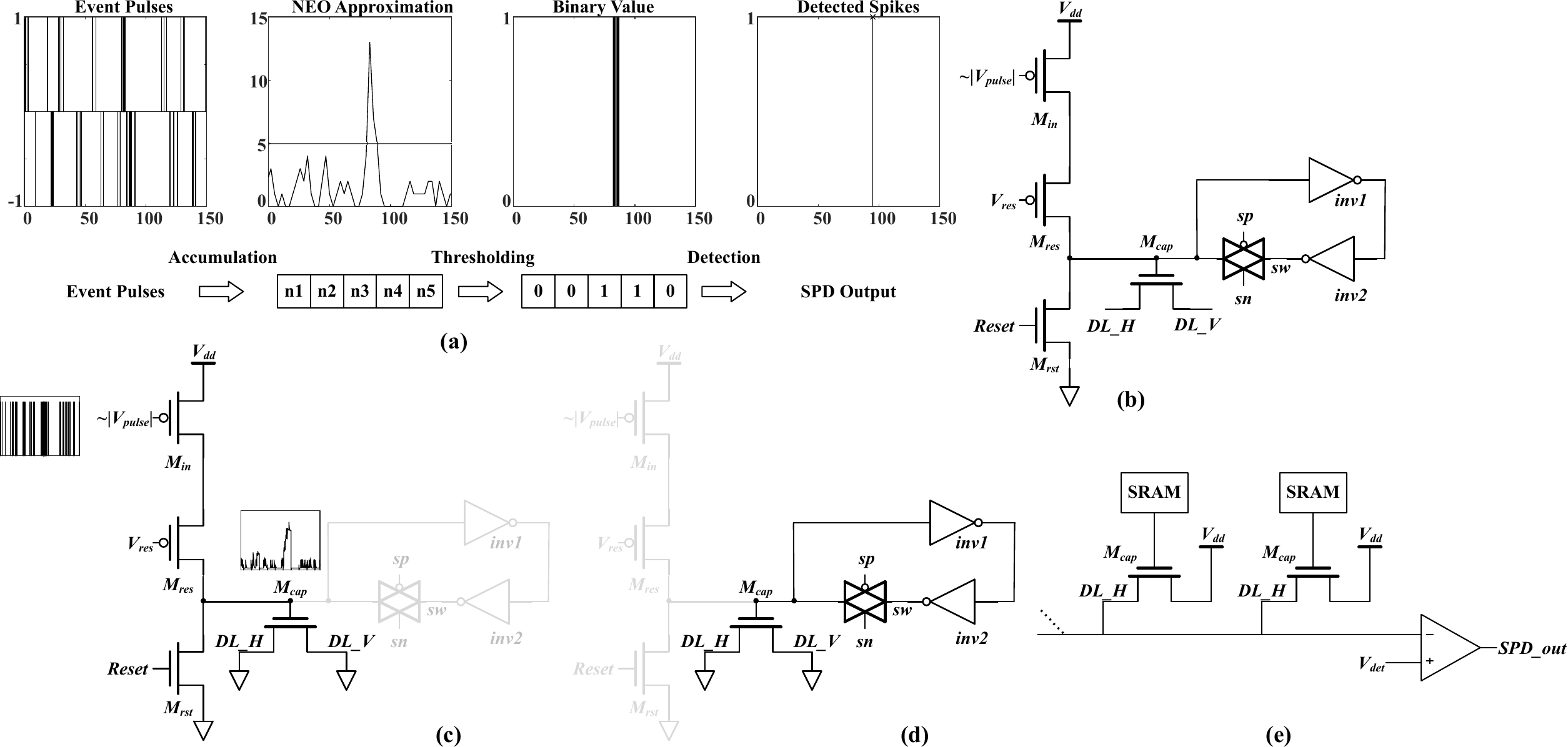}
}
\caption{
 (a) Spike detection method from event pulses based on Non-linear Energy Operator (NEO) approximation, X-axis represents time in samples. (b) 
 Proposed 10-T HRAM bitcell structure. HRAM bitcell working in (c) accumulation phase with an indication of the pulse inputs and the sum of accumulated DRAM voltages, (d) SRAM phase with first thresholding of capacitor voltage and (e) final detection phase with second thresholding.
}
\label{fig:Fig2}
\end{figure*}

Following the compressive neural frontend, significant data rate reduction could be achieved by edge computing, i.e. integrating more computation such as spike detection (SPD) at the sensor interface \cite{chattopadhyay_big_2017}. Previous studies have been reported on spike detection algorithms and hardware trade-offs \cite{zhang_adaptive_2021}\cite{malik_automatic_2016}. Some low-power neural spike detector implementations have also been proposed \cite{yao_07_2016}\cite{koutsos_15_2013}. However, earlier works have not addressed the memory requirement for storing the digital output for thousands of channels before SPD--this becomes an emergent issue with large memories\cite{horowitz_11_2014}. For example, assuming $N_{chan}=10K$, $12$-bits, $30$ kbps ADC and $n_p=7$ samples for SPD, a conventional architecture in Fig.\ref{fig:Fig1} (a) requires $102.5$ kB memory for input buffer of SPD resulting in read-dominated power of $\approx 9-24$ mW for SPD alone based on $45$ nm estimates \cite{horowitz_11_2014}. Since the amplifier array can be designed within a power budget of $10$ mW \cite{dong_han_045v_2013}, the SPD memory buffer becomes a bottleneck (latency considerations are even more demanding). While EBNR frontends can reduce writes to memory, they cannot reduce the size of the memory buffer due to unpredictability of spike locations in the buffer. Moreover, signal reconstruction and on-chip storage are still required for current solutions.

To address these challenges, we propose an In-memory computing (IMC) based SPD architecture for EBNR frontends that can perform SPD operations within memory as indicated in Fig. \ref{fig:Fig1} (b) and thus saving memory access energy \cite{bose_389_2023}.
The main contribution of this work could be summarized as follows:
\begin{enumerate}
    \item We presented and evaluated an efficient SPD method for EBNR frontends by estimating the nonlinear energy operator from the event pulses without signal reconstruction.
    \item Inspired by the hybrid SRAM-DRAM IMC for image processing in \cite{zhang_9151220_2023}, we proposed a 10T hybrid RAM (HRAM) bitcell to perform the pulse accumulation and thresholding operations for spike detection using IMC. 
\end{enumerate}

\section{Methodology}
\subsection{Spike Detection from Events}
The temporal sparsity of action potentials in neural recordings enables high compression rates when using an AER-based delta modulation mechanism. Functionally, the delta modulator produces two types of digital pulses (ON or OFF) if the input signal changes by a fixed positive or negative amount respectively. This process converts the analog spike recording $V_{in}(t)$ into an asynchronous event pulse stream $V_{pulse}(t)$. Instead of recovering a stair-step reconstruction $\hat{V}_{in}(t)=\int V_{pulse}(t)dt \approx V_{in}(t)$  as in \cite{corradi_neuromorphic_2015,mohan_architectural_2023}, we proposed a nonlinear energy operator (NEO) based spike detection method from event pulses directly, as described in Fig. \ref{fig:Fig2} (a).

Due to the noisy nature of neural recordings and the variability in spike shapes, a typical spike detection process includes an emphasizer to estimate the instantaneous change in amplitude and frequency and then applies thresholding to derive the detection output. The NEO is a commonly used emphasizer given by $\operatorname{NEO}(V_{in}(t))=\left(\frac{d V_{in}(t)}{d t}\right)^2-\frac{d^2 V_{in}(t)}{d t^2} \cdot V_{in}(t)$.
To perform SPD on event pulses, we begin with an approximation of NEO introduced in \cite{yao_07_2016}-- a low-pass filtered version of $({d V_{in}}/{d t})^2$ which can be approximated by $\int({d V_{in}}/{d t})^2 dt$. Inspired by this, we propose an approximation as follows:
\begin{equation}
\label{eq:NEO_approx}
\operatorname{NEO'}(V_{in}(t))=\int_0^{T_s} ({d \hat{V}_{in}}/{d t})^2 dt
\end{equation}
The benefit of this approximation is that ${(d \hat{V}_{in}}/{d t})$ is exactly equal to the ON/OFF pulses $V_{pulse}(t)$ from the EBNR frontend and since pulse amplitude is fixed, squaring is a trivial operation requiring only the multiplexing of both ON and OFF pulses to the same wire. While mathematically $({d \hat{V}_{in}}/{d t})^2\neq ({d V_{in}}/{d t})^2$, the approximation can still be used to distinguish neural AP from background noise with proper choice of $T_s$. For the EBNR frontend, the integral in Eq. \ref{eq:NEO_approx} reduces to a summation and the final NEO' value after $T_s$ is thresholded to indicate the presence of a neural spike in that time bin. To reduce false positives further, the binary values from $n_s$ successive time bins are summed and compared with a second threshold and a 1ms refractory period is added to give the final spike detection result. The whole process is illustrated in Fig. \ref{fig:Fig2}(a).

\subsection{Proposed HRAM Bitcell}
Inspired by an IMC-based image processor for image reconstruction, we proposed a DRAM and SRAM hybrid bitcell to implement the pulse accumulation and thresholding operations described above. Each bitcell could be used for the NEO approximation and thresholding of a time bin in the above method. Figure \ref{fig:Fig2} (b) shows the details of the proposed 10T HRAM bitcell. The bitcell can be divided into two parts: 1) a 3T1C DRAM consisting of an input access transistor $M_{in}$, a MOS capacitor $M_{cap}$, together with a pull-up variable resistor $M_{res}$ and a pull-down reset transistor $M_{rst}$; 2) An SRAM latch ($inv1,inv2$) with a transmission gate $sw$ between the output node of $inv2$ and the storage node to disable the latch when needed. In addition, a horizontal detection line $DL\_H$ connects the bitcells in a channel, and both $DL\_H$ and $DL\_V$ can be configured as pull-up, pull-down, or floating at different phases. In particular, the DRAM storage device $M_{cap}$ is a versatile component that can be used either as a MOSCAP with its drain and source grounded or as a transistor with its gate controlled by the SRAM.

The in-memory computing SPD has three sequential operation phases during each spike detection cycle: accumulation phase, thresholding phase, and detection phase.
\subsubsection{Accumulation Phase}
In this phase, the SRAM latch is disabled by setting $sp$ = 1 and $sn$ = 0 to give the equivalent circuit in Fig. \ref{fig:Fig2} (c).  Both detection lines, $DL\_H$ and $DL\_V$, are connected to the ground, which allows the use of $M_{cap}$ as a MOSCAP.  Additionally, the active low event pulses input to the bitcell is enabled during the allocated time bin to accumulate charge on the DRAM capacitor. Each event pulse input would result in a voltage jump on $M_{cap}$, and the accumulated voltage of all input pulses over $T_s$ approximates NEO'. $M_{res}$ and $M_{cap}$ determine the voltage jump per event pulse.

\subsubsection{Thresholding Phase}
Shortly after the accumulation phase, the switch is enabled by setting $sp$ = 0 and $sn$ = 1 to latch the accumulated voltage as in Fig. \ref{fig:Fig2} (d). The trip point of the SRAM, $TH_{SRAM}$, is designed to apply a thresholding process on the NEO'. Therefore, the accumulated voltage is converted into a binary value stored in the SRAM indicating the presence of a neural AP. The SRAM holds its value until all the successive cells accomplish the accumulation and thresholding operations.

\subsubsection{Detection Phase}
After the SRAMs have latched a binary value for the time bin, $DL\_V$ is pulled up to VDD while the horizontal detection line $DL\_H$ is floating. The data stored in the SRAM determines whether the bitcell would charge the detection line as indicated in the equivalent circuit Fig. \ref{fig:Fig2} (e). The sensed potential at the detection line (corresponding to sum of $n_s$ HRAM cells in the row) is then compared with a reference detection threshold voltage to give the final spike detection output. At the end of each detection cycle, the reset signal is pulled up to clear the oldest SRAM value in the row.

\subsection{System Level Architecture}
A $N_{chan}\times n_s$ HRAM memory array is required to store the output of the EBNR frontend. The $n_s$ bitcells in every row are organized as a circular buffer by using a pointer in the memory controller. In every $T_s$ time bin, the memory controller directs events from the frontend to a column location based on this pointer. After the final detection by summing over $n_s$ cells in the row, the memory pointer is incremented and the new location is reset before a new accumulation phase starts.

\section{Results}

\begin{figure}[b!]
\centering
\resizebox{0.45\textwidth}{!}{
\includegraphics[width=\textwidth]{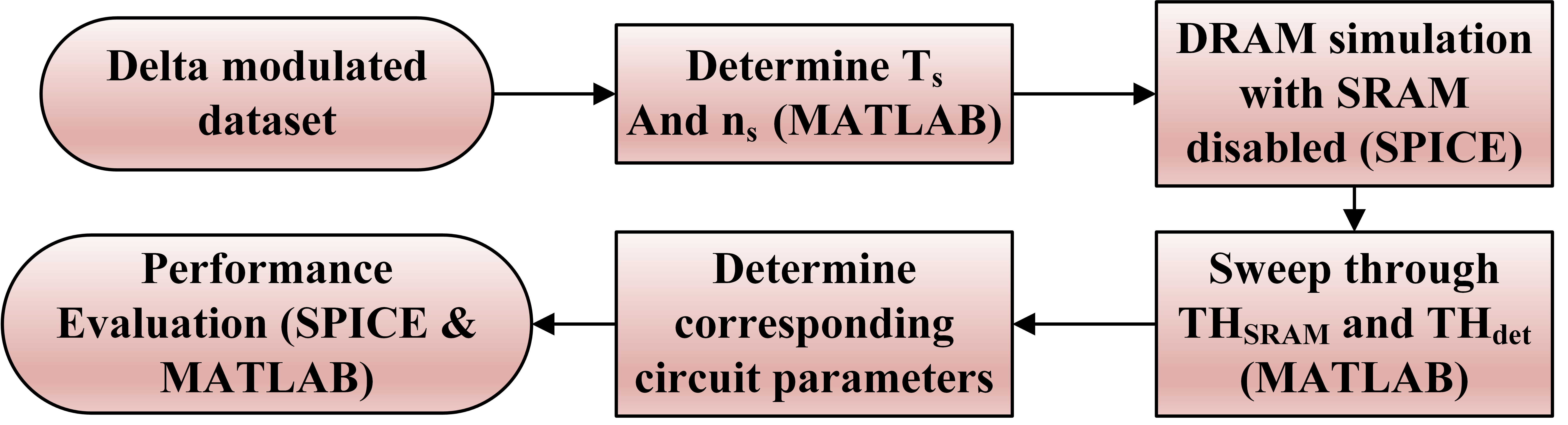}
}
\caption{
Flow chart of the data processing pipeline for HRAM-based in-memory computing spike detection simulations.
}
\label{fig:Fig4}
\end{figure}

\begin{figure}[t!]
\centering
\resizebox{0.45\textwidth}{!}{
\includegraphics[width=\textwidth]{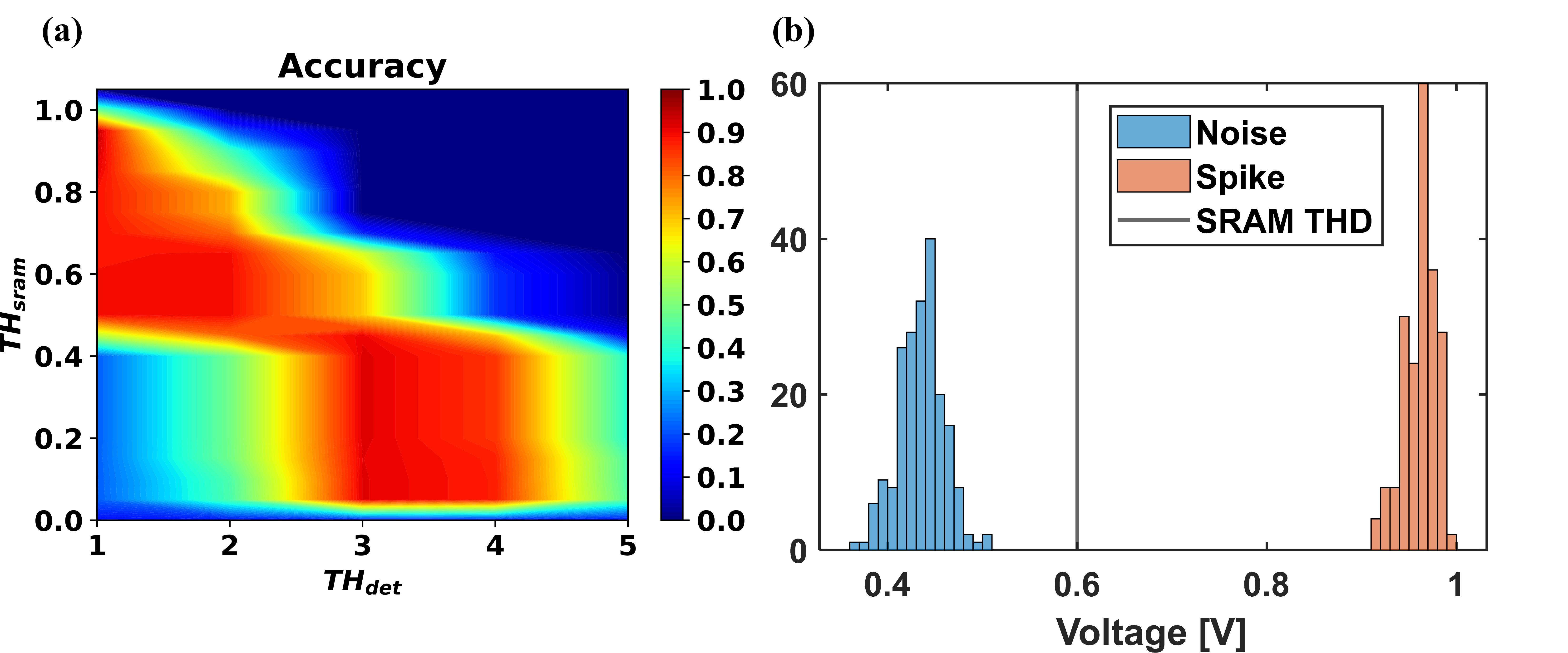}
}
\caption{
(a) Spike detection accuracy heatmap for different SRAM trip voltage and comparator threshold. (b) 200-run MC simulation results showing the variation of DRAM peak voltage for 10 ms noise recording input and action potential input.
}
\label{fig:Fig5}
\end{figure}

\begin{figure}[t!]
\centering
\resizebox{0.45\textwidth}{!}{
\includegraphics[width=\textwidth]{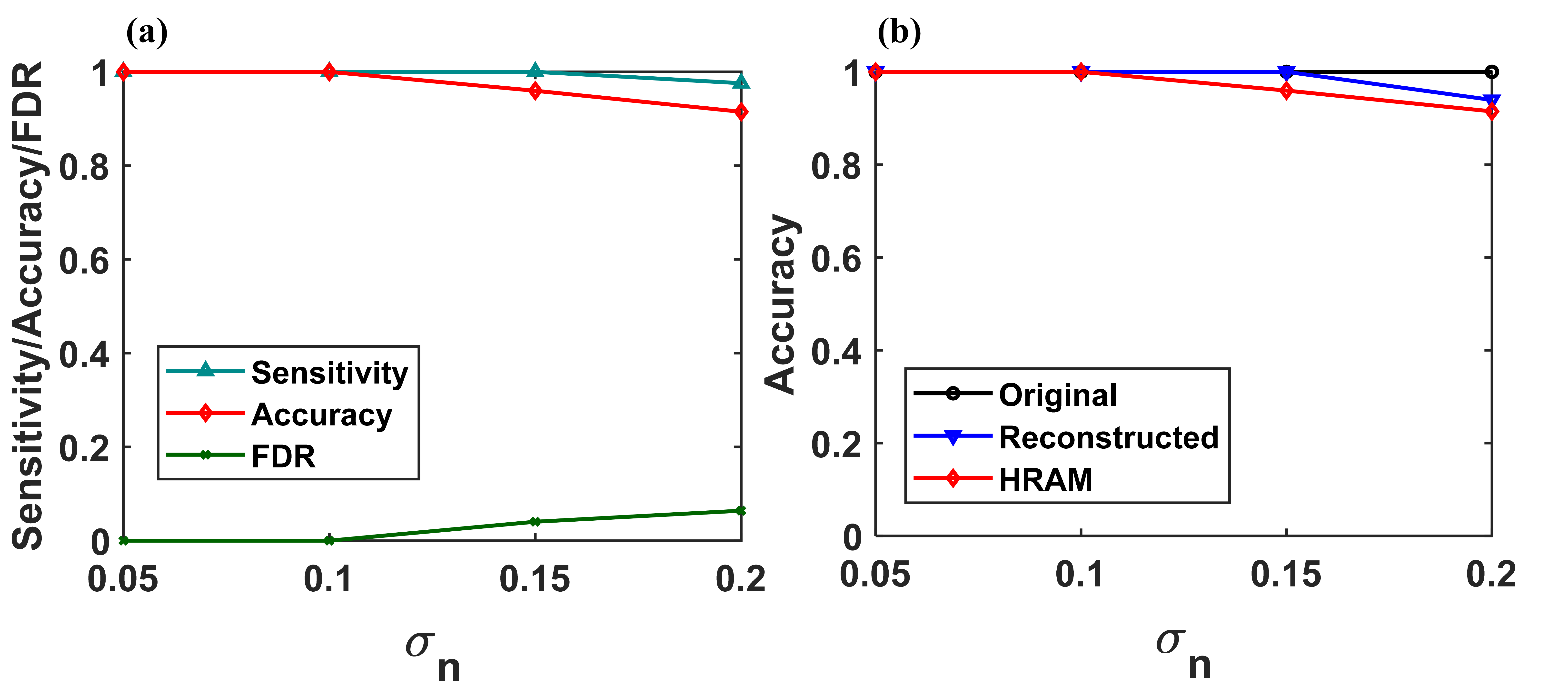}
}
\caption{
(a) Measured sensitivity, FDR and accuracy of proposed HRAM spike detection cell. (b) Comparison between proposed HRAM SPD accuracy and NEO on original signal and reconstructed signal.
}
\label{fig:Fig6}

\end{figure}
We simulated one channel of the HRAM memory array in a $65$nm CMOS process to estimate hardware performance in terms of power dissipation as well as SPD accuracy compared to software baselines. In our simulations, we followed the processing pipeline as shown in Fig.\ref{fig:Fig4}, where a SPICE-informed MATLAB model was used for design space exploration. 

\subsection{Datasets}
The dataset used in this work is a synthetic dataset provided in \cite{quiroga_unsupervised_2004}. It consists of actual spike shapes placed in time using a Poisson distribution superimposed with realistic noise at levels of 0.05, 0.1, 0.15, and 0.2. We applied delta modulation on the dataset to convert the $24$ kHz sampled neural signal into $1$ ns ON/OFF pulses similar to \cite{mohan_architectural_2023}. The event pulses were then used as input to the IMC-based spike detector.

\subsection{Choice of Parameters}
We determined optimal values of the time bin for accumulation $T_s$ and number of cells $n_s$ by running an equivalent algorithm in MATLAB. Our evaluation revealed that when the time bin is too short, it leads to a small peak of the pulse count that is hard to recognize. On the other hand, when the time bin is too long, the noise in neural signals tends to accumulate. Accordingly, $T_s\approx 125$ $\mu s$ and $n_s\approx5$ was chosen. However, as shown later, the performance is not sensitive to small changes in these values.

The choice of SRAM trip voltage ($TH_{SRAM}$) and the threshold for the detection output comparator ($TH_{det}$) are crucial parameters for IMC-based neural SPD. To optimize these parameters and validate the robustness of our HRAM-based spike detection, we ran a DRAM simulation with SRAM disabled (accumulation phase). We recorded the maximum accumulated voltage on the DRAM for each time bin using four 6-second neural recordings at different noise levels and then swept across different $TH_{SRAM}$ and $TH_{det}$ values. We averaged the accuracy over 4 noise levels to create the heatmap shown in Fig. \ref{fig:Fig5} (a). The results indicate that an SRAM binarization threshold of $\approx 600$ mV and a spike detection threshold of $2$ are optimal to reduce false detections. Moreover, the accuracy is high over a broad range of values making the detection tolerant to variations.

In order to further assess the effect of mismatch, we evaluated the process variation of the DRAM peak voltage for a $10$ ms noise input and for a $10$ ms neural spike input in a 200-run Monte Carlo simulation. We selected the size of the SRAM inverter transistors for a nominal threshold voltage of $600$ mV to distinguish between noise and neural signals. It can be seen in Fig. \ref{fig:Fig5} (b) that even with mismatch of $M_{res}$ and $M_{cap}$, the HRAM cell has a wide margin to differentiate between noise and neural spike. In addition, a negative feedback mechanism for gate voltage control in \cite{zhang_9151220_2023} can be employed to keep the MOS resistance stable with process and temperature variation.

\subsection{Measurement Results}
To validate the function of our proposed design, we tested the accuracy of spike detection. The metrics used to compare spike detection algorithms are sensitivity (S), accuracy (A) and false detection rate (FDR), as presented below : 
\begin{equation}
\text {S}=\frac{\text {TP}}{\text {TP+FN}}, 
\text {A}=\frac{\text {TP}}{\text { TP+FP+FN}}, 
\text {FDR}=\frac{\text {FP}}{\text {TP+FP}} 
\end{equation}
where True Positive (TP) represents truly detected spikes, False Positive (FP) represents wrongly detected spikes while False Negative (FN) represents the undetected spikes.
The single-channel IMC spike detector takes 6-second delta-modulated neural recordings at four noise levels as input. Fig. \ref{fig:Fig6} (a) shows the spike detection accuracy at different noise levels. The X-axis of the plot  $\sigma_{n}$ is the noise level normalized to the spike amplitude. It can be seen that the proposed HRAM-based spike detector has a high accuracy for low-noise neural input. At high noise level, the proposed method could still retain relatively high sensitivity but a drop in accuracy is observed due to FDR rise.

In Fig. \ref{fig:Fig6} (b) we compared the proposed HRAM-based spike detector with software NEO-based spike detection with a threshold multiplier setting in \cite{oprea_hardware_2022} applied both on the original neural signal and the stair-step reconstructed signal from event pulses similar to \cite{mohan_architectural_2023}. The proposed HRAM-based spike detector outperformed the reconstructed signal at low noise since the proposed method had higher sensitivity while the reconstructed signal is affected by the reconstruction noise, but the accuracy of the proposed method at high noise level is slightly lower than software NEO at around 92\%.

We have compared our proposed low-power neural spike detector with other recently proposed detectors in Table \ref{Table:Comparison}. Our HRAM bitcells consume an estimated power of 0.32 nW per channel, excluding the peripheral circuitry and comparators. Other implementations may consume more power due to the power consumed by memory banks. However, our design maintains comparable detection accuracy while achieving very low power consumption. Additionally, the memory requirement for the SPD buffer (Fig. \ref{fig:Fig1}) has been reduced to only $N_{chan}\times n_s = 6.1$ kB from $102.5$ kB for the conventional approach. Despite the $\approx 4X$ increase in bitcell size of HRAM compared to foundry SRAM, the proposed system achieves a $4.2X$ reduction in memory area.

\begin{table}[t]
\caption{Comparison with other low-power neural spike detectors}
\label{Table:Comparison}
\centering
\label{tab}
\resizebox{0.49\textwidth}{!}{
\begin{tabular}{lcccc}
 \toprule
 & This Work         & \cite{yao_07_2016}    & \cite{zhang_calibration-free_2023}   & \cite{guo_accurate_2022}   \\
 \midrule
Technology (nm)           & 65                & 65     & 65      & 65       \\
Implementation                     & Analog            & Analog & Digital & Digital  \\
Supply Voltage (V)        & 1.0                 & 0.7    & 1.2     & 1.1      \\
Pre-emphasizer            & Event NEO & ED+LPF & ADF     & Dual NEO \\
Power (nW/Ch) & 13.8            & 40     & 38      & 70       \\

SPD technique            & IMC             & Analog & Digital & Digital  \\
Accuracy [min,max]               & [92\%,99\%]              & [94\%,99\%]   & [96\%,99\%]    & [97\%,99\%]    \\
\bottomrule

\end{tabular}
}
\end{table}

\section{Conclusion}
In this work, we proposed an efficient HRAM IMC-based spike detector for compressive EBNR frontend. We proposed a SPD method from event pulses without signal reconstruction by accumulating event pulses and thresholding for a binary value indicating the pulse density. We also designed a 10-T SRAM and DRAM hybrid in-memory computing bitcell for accumulation and thresholding operations. We simulated the single-channel spike detector and showed that our method achieves SPD accuracy $\approx92-99\%$ while dissipating only $13.8$ nW per channel and enabling $4.2X$ reduction in memory area.

\clearpage
\bibliographystyle{IEEEtran}
\bibliography{references}

\end{document}